\documentclass[prd,twocolumn,nofootinbib,amsmath]{revtex4}

\usepackage{amsmath,amssymb}
\usepackage{graphicx,units}
\usepackage{hyperref}

\begin{document}

\title{Beta Distribution of Human MTL Neuron Sparsity: A Sparse and Skewed Code}

\author{Andrew Magyar}
\affiliation{Physics Department,
   Pennsylvania State University,
   University Park, PA 16802, USA
}
\date{08 Aug 2015}

%=======================

\begin{abstract}
Single unit recordings in the human medial temporal lobe (MTL) have revealed a population of cells with conceptually based, highly selective activity, indicating the presence of a sparse neural code. Building off previous work by the author and J.C. Collins, this paper develops a statistical model for analyzing this data, based on maximum likelihood analysis. The goal is to infer the underlying distribution of neural response probabilities across the population of MTL cells. The response probability, or neuronal sparsity, is defined as the total probability that the neuron produces an above-threshold firing rate during the presentation of a randomly selected stimulus. 
Applying the method, it is shown that a beta-distributed neuronal sparsity across the cells of the MTL is consistent with the data. The resulting fits reveal a sparse and highly skewed code, with a huge majority of neurons exhibiting extremely low response probabilities, and a smaller minority possessing considerably higher response probabilities. The distributions are closely approximated by a power law at low sparsity values. Strikingly similar skewed distributions have been found in the statistics of place cell activity in rats, suggesting similar underlying coding dynamics between the human MTL and the rat hippocampus.
\end{abstract}

\maketitle

%=======================

\section{Introduction}
\label{sec:intro}
%Sparseness has been identified as a prevalent feature of neural representations of external stimuli across multiple sensory modalities [REFERENCES: Rolls Tovee 1995, DeWeese 2003, Perez-Orive 2002), OTHERS?] as well as in limbic structures [REFERENCES: \cite{HVC-RAX}, \cite{GMC}, OTHERS?]. Specifically, the neural code has characteristics of both "`population sparseness"' and "`lifetime sparseness,"' where activity is evoked in a relatively small fraction of neurons during a stimulus presentation, and where neurons are responsive to only a very small fraction of stimuli, respectively [REF: Willmore Tolhurst 2001]. 

%Attempts to characterize sparseness have resulted in the proliferation of several quantitative measures [REFERENCES: \cite{ison}, Quian Quiroga decoding 2007], based upon either neuronal firing rate distribution [REFERENCE: Rolls Tovee 1995, Vinje Gallant 2000, Leahy Sejnowski 2005] or upon a binary threshold, in which a neuron is treated as either silent or responsive during stimulus presentation [REFERENCE: \cite{Mormann}, Waydo 2006, Quian Quiroga 2007]. It has been shown that the commonly-used measure for lifetime sparseness introduced by Rolls and Tovee [REFERENCE: Rolls Tovee 1995] is insensitive to the responsive properties of extremely sparse neurons [References: Quian Quiroga 2007, Magyar Collins], where neurons produce highly elevated firing rates for only a very small proportion of stimuli. In such cases, the binary threshold measures are preferred. 

The sparse coding hypothesis states that neural processing of sensory information is organized to produce representations of salient aspects of the environment (people, objects, landmarks, etc.) using only a small number of strongly activated neurons \cite{barlow}. This kind of representation lies between two theoretical extremes: dense coding, in which each stimulus is represented in the activation of a substantial proportion of the available cells; and local coding, where each object is represented by the firing of a single neuron \cite{foldiak1,foldiak2,bowers}. Sparse coding schemes possess many favorable properties including a high storage capacity, energy efficiency, %\cite{Levy, Attwell, Lennie}
 and ease of readability (for references and further discussion, see the review by Olshausen and Field \cite{olshausen.sparse}).

Experimental detection of sparse codes involves identifying numerous cells that are highly selective, responding to a very small proportion of complex stimuli \cite{willmore2001, willmore2011}. Examples include odor-specific Kenyon cells in locusts \cite{perez}, V1 cells in cats \cite{baddeley} and mice \cite{froudarakis}, and neurons in the temporal cortex in non-human primates \cite{baddeley, rolls1995, rust, vinje}. Also, the RA projecting neurons of the HVC in zebra finches exhibit extremely sparse activity during song production \cite{hahnloser}. In humans, striking evidence of sparse coding has been observed in the medial temporal lobe in a series of experiments \cite{fried, kreiman, qq1, qq.multi} including the concept cells reported in Quian Quiroga et al. \cite{qq1} which were observed to respond to stimuli related to a single concept (e.g. the ``Jennifer Aniston neuron'')  out of nearly $100$ other concepts presented.

Therefore, characterizing the sparseness of the neuronal representations is an important goal. Towards this end, one metric used is the total fraction of stimuli that elicit a response in a particular neuron, that we term the neuronal sparsity, $\alpha$ \cite{twopop}. 
\begin{equation}
\alpha=\frac{\text{\# stimuli triggering a response}}{\text{total \# of possible stimuli}}
\label{alpha}
\end{equation}
This definition of sparsity is a property of the neuron itself, and is the total probability that the cell responds to a randomly chosen stimulus. It is not necessarily equal to the fraction of stimuli that elicit a response during a particular experiment in which only a small subset of possible stimuli are presented, as used in Ison et al. \cite{ison} for example. If a particular cell remains unresponsive during the presentation of $100$ randomly selected images, then its sparsity is not necessarily $0$, it may instead have a very low but non-zero sparsity. In other words, $\alpha$ is a quantity that must be inferred from a particular experiment rather than directly calculated. %Analyses of the data reported in \cite{GMC} found that for most neurons,  $\alpha < 0.03$ \cite{Waydo, Ison}. Recent work on the data reported in \cite{Mormann} has shown that sparsity is highly variable throughout the neurons of the human MTL \cite{CJ,twopop}, with most neurons ($\approx95\%$) responding to a small proportion ($\alpha\approx10^{-3}$) of stimuli while the rest respond more frequently ($\alpha\approx10^{-2}$).

This approach treats neuronal activity as binary ``active-vs-inactive'' over a certain post-stimulus time window. This is appropriate for cells that exhibit highly elevated firing rates under specific circumstances compared with the baseline rate \cite{qq.decoding, twopop}, and few in-between cases. Place cells, for example, display high firing rates when the organism is at a particular location in the environment and much lower firing rates otherwise \cite{okeefe, wilson}.

The principal goal of this paper is to extend the model developed by the author and Collins \cite{twopop} for fitting the human MTL data presented in Mormann et al. \cite{mormann} in order to estimate the distribution of sparsity across the population of neurons. In that work, the cells were split into two discrete populations, each with a characteristic sparsity value. In this paper, the model is extended to include continuous sparsity distributions. 

Specifically, the neuronal sparsity in the MTL is assumed to follow a beta distribution. Motivation for choosing the beta distribution comes from its common usage as a distribution of probabilities, giving it wide application in Bayesian statistics \cite{mackay}. The PDF with parameters $a$ and $b$ is given by
\begin{equation}
D_{a,b}\left( \alpha \right)=\frac{\alpha^{a-1} \left(1-\alpha \right)^{b-1}}{B\left(a,b\right)}
\label{beta}
\end{equation}
where the normalization factor $B\left(a,b\right)$ is the beta function. This produces an overdispersed model for the neural responses, predicting a heavier tail than if all neurons had the same sparsity. Such overdispersed, or skewed models, are suspected to play a prevalent role in many aspects of neural systems \cite{buzsaki.2014,buzsaki.2015}, and such distributions have been found in place cell activity of the rat CA1 \cite{rich}. 

%Perhaps the two most striking examples of sparse codes are the place cells found in the rodent hippocampus \cite{okeefe, } In a recent set of experiments, single-unit recordings in humans have found such ultra-sparse, highly selective neurons in the medial temporal lobe (MTL) \cite{fried, krieman, qq1, qq.multimodal}. Among these include the class of cells referred to as ``Jennifer Aniston neurons,'' or concept cells, reported in \cite{qq1}, which responded exclusively to images of a single celebrity and remained largely silent during the presentation of images of roughly $100$ other familiar celebrities and landmarks. 

%It has been suggested that these neurons are concept cells, responding to stimuli associated with a particular concept (or a small set of concepts), and that they play a key role in declarative memory functions and cognition \cite{qq1}. 

%Taking into consideration the favorable theoretic properties of sparse coding, the diverse experimental evidence, and the important role it plays in human cognition, sparse coding has been proposed as a fundamental aspect of sensory processing \cite{olshausen.sparse, qq.concept}. 
  
Following the fitting procedure developed by the author and Collins in \cite{twopop}, we perform inferential statistics, treating both the neurons and the presented stimuli as random samples from their respective ``universe''. This way, we can infer the response properties of all neurons in a brain region, rather than merely describing the data, as is done in previous analyses \cite{ison}. The data reported in Mormann, et al. \cite{mormann} is fit to the beta model using maximum likelihood analysis, yielding the best estimates of the beta distribution parameters, $a$ and $b$. Then, the goodness of fit is assessed using $\chi^2$ analysis. 

The model is shown to produce acceptable fits in all four subdivisions of the MTL for which data is available: the hippocampus (Hipp), the entorhinal cortex (EH), the amygdala (Amy), and the parahippocampal cortex (PHC). The estimated beta distributions for the population of human MTL neurons indicate:
\begin{itemize}
	\item The neuronal sparsity is highly non-uniform, with the least selective $5\%$ of neurons possessing a sparsity over $0.01$.
	\item $D_{a,b}\left( \alpha \right)$ nearly has a power-law divergence with exponent $\approx-1$ at low sparsity.
	\item The mean sparsity of MTL neurons is very low, on the order of $10^{-3}$.
\end{itemize}
This model marks an improvement over the model previously developed in \cite{twopop}, where the neurons were assumed to be split into two populations, each characterized by a sparsity value. This model was able to produce good fits in the Hipp and EC, but it failed to fit the data from Amy and PHC. Furthermore, the beta model has only two fitting parameters, while the two-population model has three. 

Finally, the skewed, highly non-uniform beta-binomial distributions of responses in the human MTL are compared with strikingly similar results recently observed in the statistics of place cell activity of the rat hippocampus reported in Rich et al. \cite{rich}. Specifically, they observe that the number of place fields recruited per cell follows a skewed gamma-poisson distribution. Gamma-poisson distributions are simply limiting cases of beta-binomial distributions (shown in Appendix \ref{appendix.gamma-poisson}). This strongly suggests that place cell codes and concept cell codes are two different manifestations of the same underlying dynamics, in line with previous suspicions \cite{qq.concept}. 

\section{Beta-Binomial Model}
\label{sec:BBM}

We define the sparsity, $\alpha$, of a particular binary neuron according to Eq. (\ref{alpha}). In the experiment we analyze, a neuron is considered responsive to a particular stimulus if its firing rate exceeds the baseline rate by a statistical threshold during an appropriate post-stimulus time frame (see \cite{mormann} for details). %This binary approximation of neural responses is appropriate for cells that have a very low baseline firing rate that is many orders of magnitude lower than the evoked rate \cite{twopop, decoding}, e.g. the place cells of the rat hippocampus \cite{lee.rao}. The neurons are assumed to be statistically independent. 

We define a stimulus as an image of a individual object, person, building, etc. as used in \cite{mormann}. The set of stimuli presented to the patients in \cite{mormann} constitute a tiny random sample from the universe of all stimuli.
 %Others [Rolls, Tovee; Vinje-Gallant] use a different definition in terms of the average neuronal firing rate during the presentation of a stimulus:
%\begin{equation}
%S=\frac{E[r]^2}{E[r^2]}
%\end{equation} 
%It has been shown [Quiroga 2007, 2POP] that for extremely sparse neurons, this quantity is dominated by the baseline firing rate and is thus ill-suited for human MTL data. Furthermore [DIFFICULT TO INTERPRET].
In this section, we extend our previous procedure to include continuous distributions of sparsity, $D_{\bar{\theta}}\left( \alpha \right)$, with fitting parameters $\bar{\theta}$. In particular, we postulate the sparseness of a randomly selected neuron from the MTL is sampled from a beta distribution with pdf given by equation (\ref{beta}). 

Examining the data presented in Mormann et al. \cite{mormann}, we note that there are a large proportion of unresponsive cells, and that a significant proportion of responsive cells respond to only one image. Hence, we would expect a sparsity distribution skewed towards zero sparsity. Note that Eq. (\ref{beta}) diverges as $\alpha \rightarrow 0$ when $a<1$, which is what we would expect for this data. For $a\approx0$, the distribution behaves like a power law as $\alpha \rightarrow 0$, however, at $a=0$ the distribution diverges too greatly at zero sparsity to be normalized.

Let $S$ be the number of stimuli presented to the patient during an experiment, and let $N$ be the number of recorded neurons, each with a sparsity sampled from \ref{beta}. The neurons are assumed statistically independent, as measured in \cite{qq1}. For a particular neuron, let $K$ equal the number of stimuli that were measured to evoke a response in that neuron, and let $n_k$ for $k=1...S$, be the number of neurons that respond to $k$ out of the $S$ stimuli presented. Then, we follow earlier work \cite{twopop} and derive the likelihood function. 

For the data we analyze, we must take into consideration that not all units isolated by the spike sorting algorithm consist of a single neuron \cite{qq1,mormann}. Limitations in the spike sorting procedure make it so that some fraction of the recorded units represent the activity of multiple neurons. If the activity of a unit is the combined activity of several neurons, then on average that unit will respond to more stimuli over the course of an experiment compared with a unit consisting of a single neuron. Thus, we will carry out the calculation in two cases: for the first case we assume all units consist of a single neuron, and for the second case we assume some fraction of units, $p$, are comprised of two neurons while the rest consist of single neurons. 

%=======================
\subsection{Derivation of Beta-Binomial Response Probability}

In this subsection, the relevant results developed by the author and Collins in \cite{twopop} are summarized, making appropriate modifications for the introduction of a continuous distribution of sparsity.

During the presentation of $S$ randomly selected stimuli, a single pseudo-binary unit with sparsity, $\alpha$, responds to $K=k$ of the stimuli with a conditional probability given by the binomial distribution (\cite{twopop}):
\begin{equation}
P\left(K=k \mid \alpha \right)
= \binom{S}{k} \, \alpha ^{k} (1-\alpha) ^{S-k}.
\label{binomial} 
\end{equation}
If the sparsity, $\alpha$ is sampled from a continuous distribution, $D_{\bar{\theta}}\left( \alpha \right)$ then, the total probability, $\epsilon_k \left( \bar{\theta} \right)$ that a randomly selected unit responds to $k$ stimuli is given by:
\begin{align}
\label{binomial.mixed}
P\left( K=k \right)
& =\binom{S}{k}\int_0^1 d\alpha \, D_{\bar{\theta}}\left( \alpha \right) 
\, \alpha ^{k} (1-\alpha) ^{S-k}  \\
& \equiv \epsilon_k \left( \bar{\theta} \right). \nonumber
\end{align}
Substituting Eq. (\ref{beta}) into Eq. (\ref{binomial.mixed}) and evaluating the integral yields the beta-binomial distribution:
\begin{equation}
		\epsilon_k \left( a, b \right)=\binom{S}{k} \frac{B\left(a+k,b+S-k\right)}{B\left(a,b\right)}.
\label{bet.bin}
\end{equation}
Eq (\ref{bet.bin}) represents the probability that a single neuron responds to $k$ out of $S$ presented stimuli. Mixing the binomial response with a beta distribution over the parameter $\alpha$, produces an overdispersed distribution of neural responses. This allows the heavy tails present in the data presented in Mormann et al. \cite{mormann}, displayed in Table (\ref{table:rawdata}). 

This allows us to fit the outcome of an experiment in which $S$ stimuli are presented to $N$ neurons, recorded in parallel. The result of the experiment is the set of numbers, $n_0$, $n_1$, $n_2$, ..., $n_S$, where $n_k$ is the number of recorded neurons that respond to $k$ of the stimuli. The expected number of cells per bin, $n_k^*$ assuming the responses are sampled from Eq. (\ref{bet.bin}), is given by:
\begin{equation}
n_k^*=N \epsilon_k \left( a, b \right) \pm \sqrt{N \epsilon_k \left( a, b \right) \left(1-\epsilon_k \left( a, b \right) \right)}.
\label{prediction}
\end{equation}
To find the values of $a$ and $b$ that provide the best $n_k^*$, we employ the method of maximum likelihood. This involves maximizing the likelihood function for the data given the beta model, $\mathcal{L}\left( a, b \right)$. The likelihood function is derived in\cite{twopop}, and the same result applies here, giving:
\begin{equation}
    \mathcal{L}\left( a, b \right) =
    W\{n_0,n_1,...,n_S\}\prod_{k=0}^{S}{[\epsilon_k \left( a, b \right)]^{n_k}}.
\label{likelihood}
\end{equation}
where the normalization constant $W\{n_0,n_1,...,n_S\}$ is the multinomial coefficient, i.e. the number of ways of rearranging $N$ objects without changing the $n_k$ values. Since $W\{n_0,n_1,...,n_S\}$ is independent of the parameters $a$ and $b$, it does not need to be taken into consideration during maximization. Maximizing $\mathcal{L}\left( a, b \right)$ gives the parameter values, $a_0$ and $b_0$ that give the best fit, $n_k^*$. This was performed numerically using Mathematica. 

The expectation values, $n_k^*=N\epsilon_k \left( a_0, b_0 \right)$, are compared with the data, $n_k$, using $\chi^2$ analysis to assess the goodness of fit. The $\chi^2$ statistic is defined:
\begin{equation}
  \label{eq:chi2}
  \chi^2(k_{\rm max}; n_1, n_2, \dots)
  = \sum_{k=1}^{k_{\rm max}} \frac{ (n_k-N\epsilon_k)^2 }{ N\epsilon_k }.
\end{equation}
For a good fit, 
\begin{equation}
\chi^2 \sim \left( \text{ \# of data points} \right) - \left( \text{\# of model parameters} \right)
\end{equation} 
This procedure only applies for $n_k^*$ larger than a few, so for our model, we fit only the first five data points ($n_0$, $n_1$, $n_2$, $n_3$, and $n_4$) in order to include only the bins with significant responses. Thus, the fit is good when $\chi^2 \sim 3$.

%=======================

\subsection{Extension of Model to Multiple-Neuron Units}

This subsection also follows the procedure developed in \cite{twopop} for incorporating the effect of multiple-neuron units. The experimentalists estimated the fraction of multiple-neuron units to be $p=0.66$ \cite{qq1}. As in \cite{twopop}, we make the simplifying assumption that all units consisting of multiple cells consist of two neurons. 

Let $\epsilon_k'\left(a,b\right)$ be the total probability that a unit selected at random responds to $k$ out of $S$ stimuli. We follow \cite{twopop} to derive the unit level response, $\epsilon_k'\left(a,b\right)$, in terms of the neuron-level response given by $\epsilon_k$. 

Then,
\begin{equation}
\epsilon_{k}'\left(a,b\right)=\left(1-p\right)\epsilon_k+p \epsilon_{(2),k}
\label{totalprob}
\end{equation}
where $\epsilon_{(2),k}$ is the probability that a randomly chosen double-neuron unit responds to $k$ stimuli. $\epsilon_{k}$ is the probability of a single neuron unit responding to $k$ stimuli and is given by equation (\ref{bet.bin}).

In order to derive $\epsilon_{(2),k}$, we first note that a double-neuron unit responds to a stimuli if either one of its constituent neurons responds. Let $\alpha_1$, $\alpha_2$ be the sparsities of the constituent neurons. Then, assuming the two neurons are independent, the unit has an effective sparsity given by
\begin{equation}
\alpha_{unit}=1-\left( 1-\alpha_1 \right) \left( 1-\alpha_2 \right)
\label{unitalpha}
\end{equation}
Similar to the single-neuron unit, the conditional probability that the double-neuron unit responds to $k$ stimuli out of $S$, assuming the unit sparsity is known, is given by the binomial distribution with probability parameter $\alpha_{unit}$. 
\begin{equation}
P\left(K=k \mid \alpha_{unit} \right)
= \binom{S}{k} \, \alpha_{unit} ^{k} (1-\alpha_{unit}) ^{S-k}.
\label{binomial.prime} 
\end{equation}
The total probability, $\epsilon_{(2),k}$ is found by integrating the conditional probability over the beta distribution, equation (\ref{beta}) for each constituent neuron:
\begin{align}
\epsilon_{(2),k}=\int_0^1{d\alpha_1}\int_0^1{ d\alpha_2} &  \big[ D_{a,b}\left( \alpha_1 \right) D_{a,b}\left( \alpha_2 \right)\, \times \nonumber\\
& \binom{S}{k} \, \alpha_{unit}^{k} (1-\alpha_{unit}) ^{S-k} \big]
\label{multiepsilon}
\end{align}
Evaluating this integral yields (derivation in appendix):
\begin{align}
\epsilon_{(2),k}= \frac{1}{[B\left(a,b\right)]^2} \,& \binom{S}{k} \, \sum_{j=0}^k  { \big\{ \, \binom{k}{j} } \,  \nonumber\\
& \times B\left(a+j , b+S-j \right) \nonumber \\
& \times B\left( a+k-j , b+S-k \right) \big\}
\label{multibbn}
\end{align}

The form of the likelihood function is the same as (\ref{likelihood}) except that Eq. (\ref{totalprob}) is used instead of Eq. (\ref{bet.bin}) for the single unit response probability. 
\begin{equation}
    \mathcal{L}'\left( a, b \right) =
    W\{n_0,n_1,...,n_S\}\prod_{k=0}^{S}{[\epsilon_k'\left(a,b\right)] ^{n_k}}.
\label{likelihood.prime}
\end{equation}
Maximizing (\ref{likelihood.prime}) with respect to $a$ and $b$ provides the best fit parameters of the model $a_0$ and $b_0$ to the $n_k$ data. The model prediction, $n^*_k$ has the same form as equation (\ref{prediction}) only with $\epsilon_k'$ in place of of $\epsilon_k$. 

%=====================

\section{Results}
\label{sec:results}

%=======================

\subsection{Fits to data}

\begin{figure*}
\centering
\begin{tabular}{c@{\hspace*{6mm}}c}
    \includegraphics[scale=0.4]{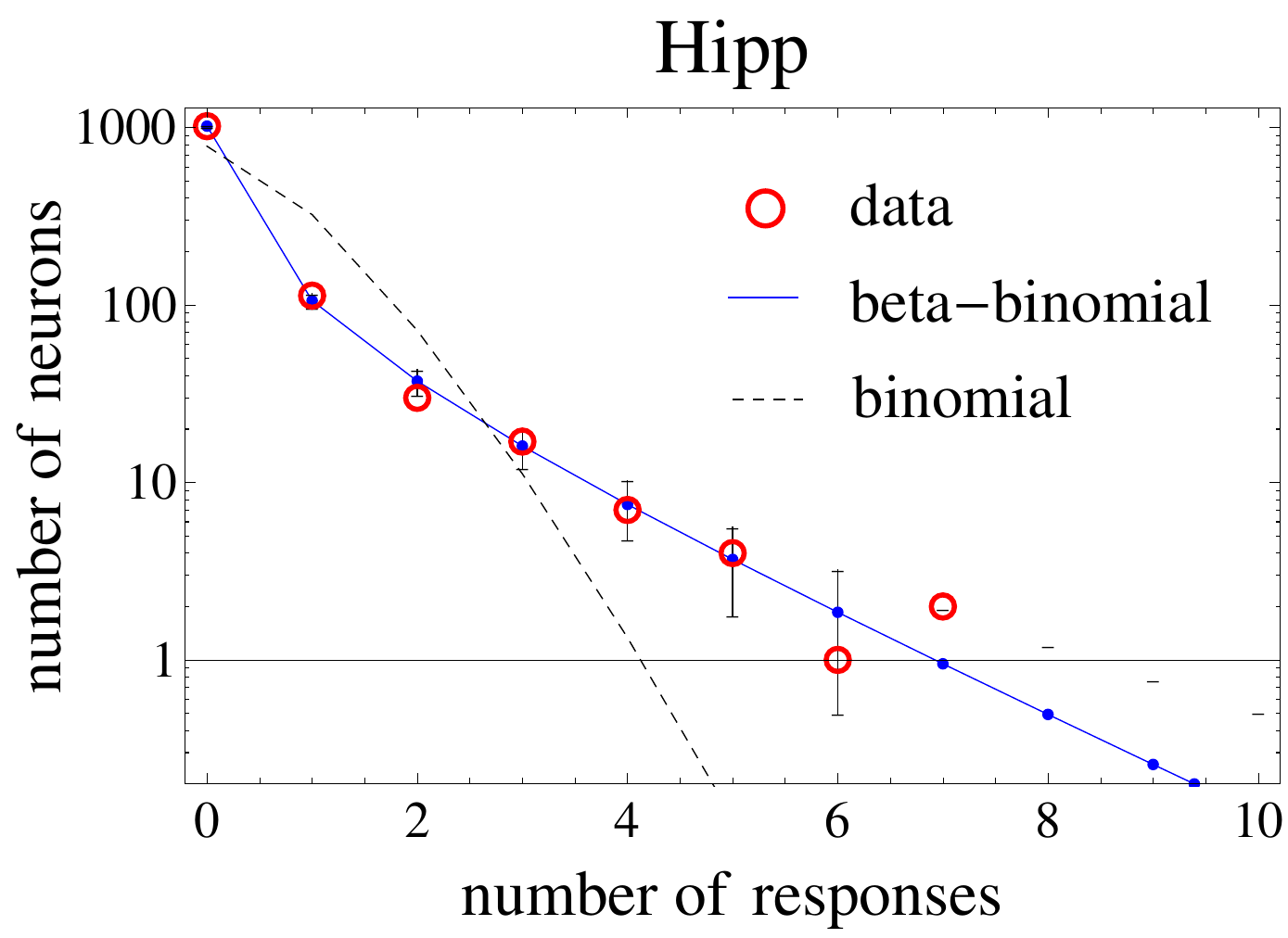} &  \includegraphics[scale=0.4]{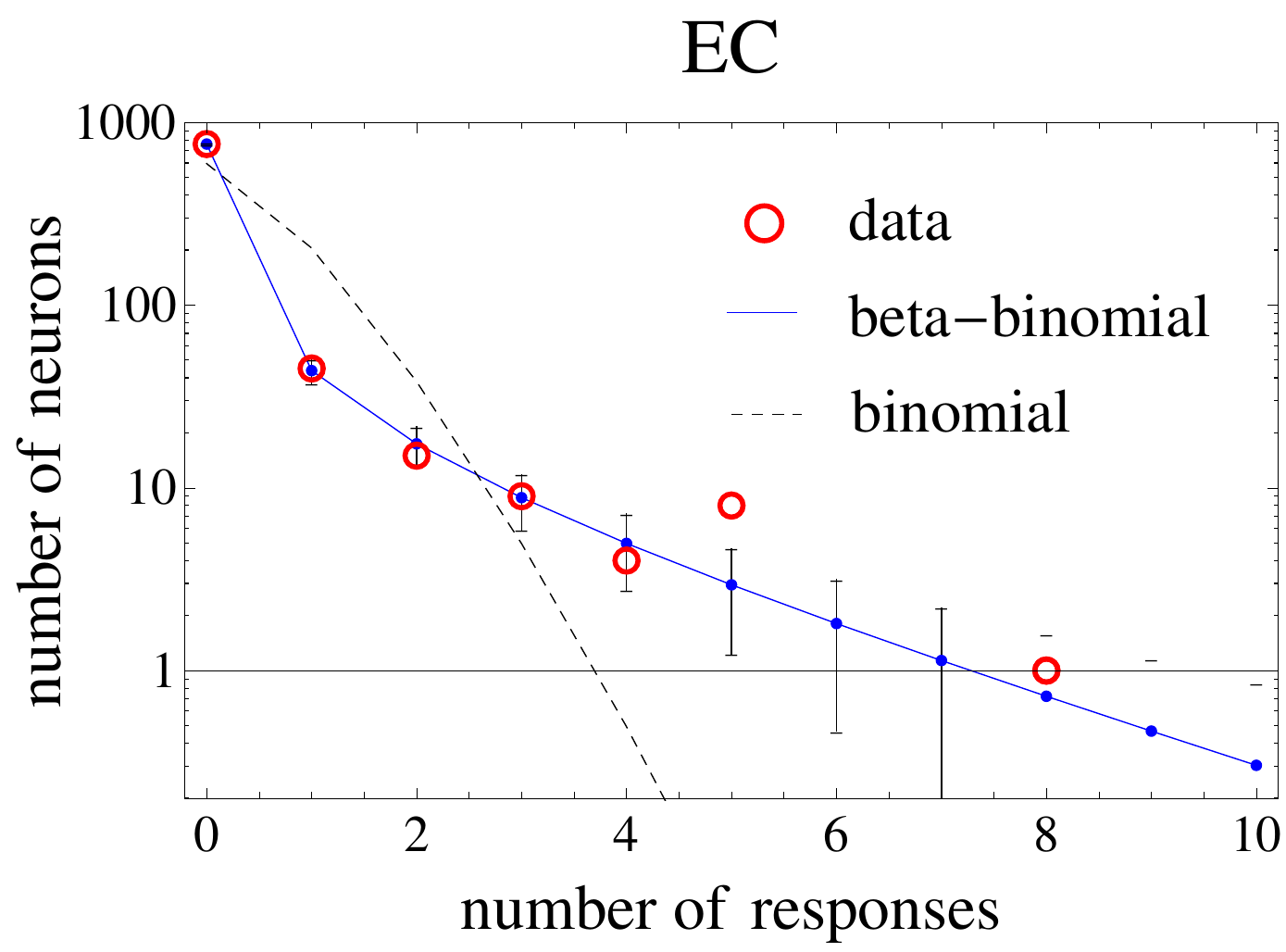} 
    \\ \includegraphics[scale=0.4]{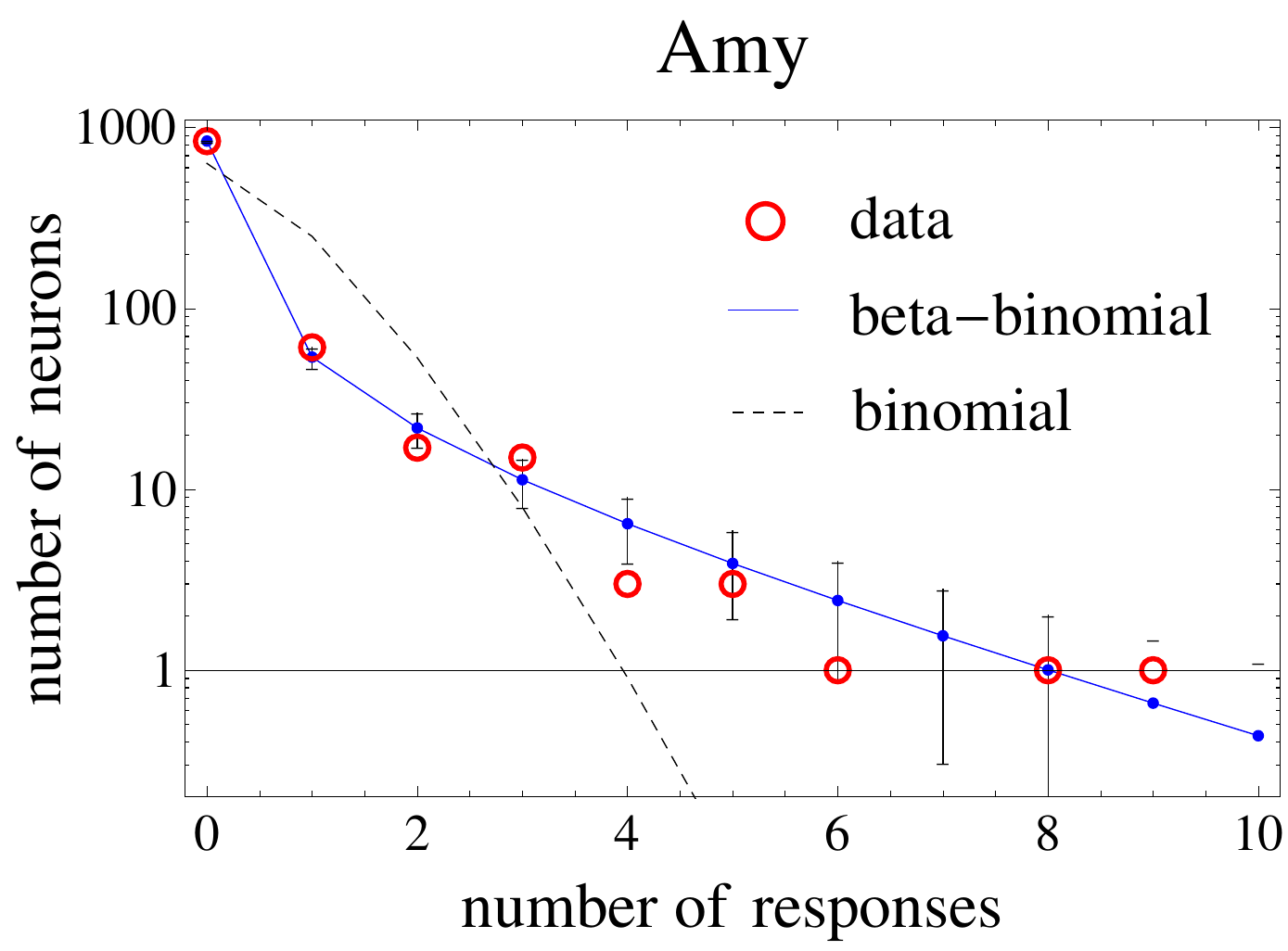} &  \includegraphics[scale=0.4]{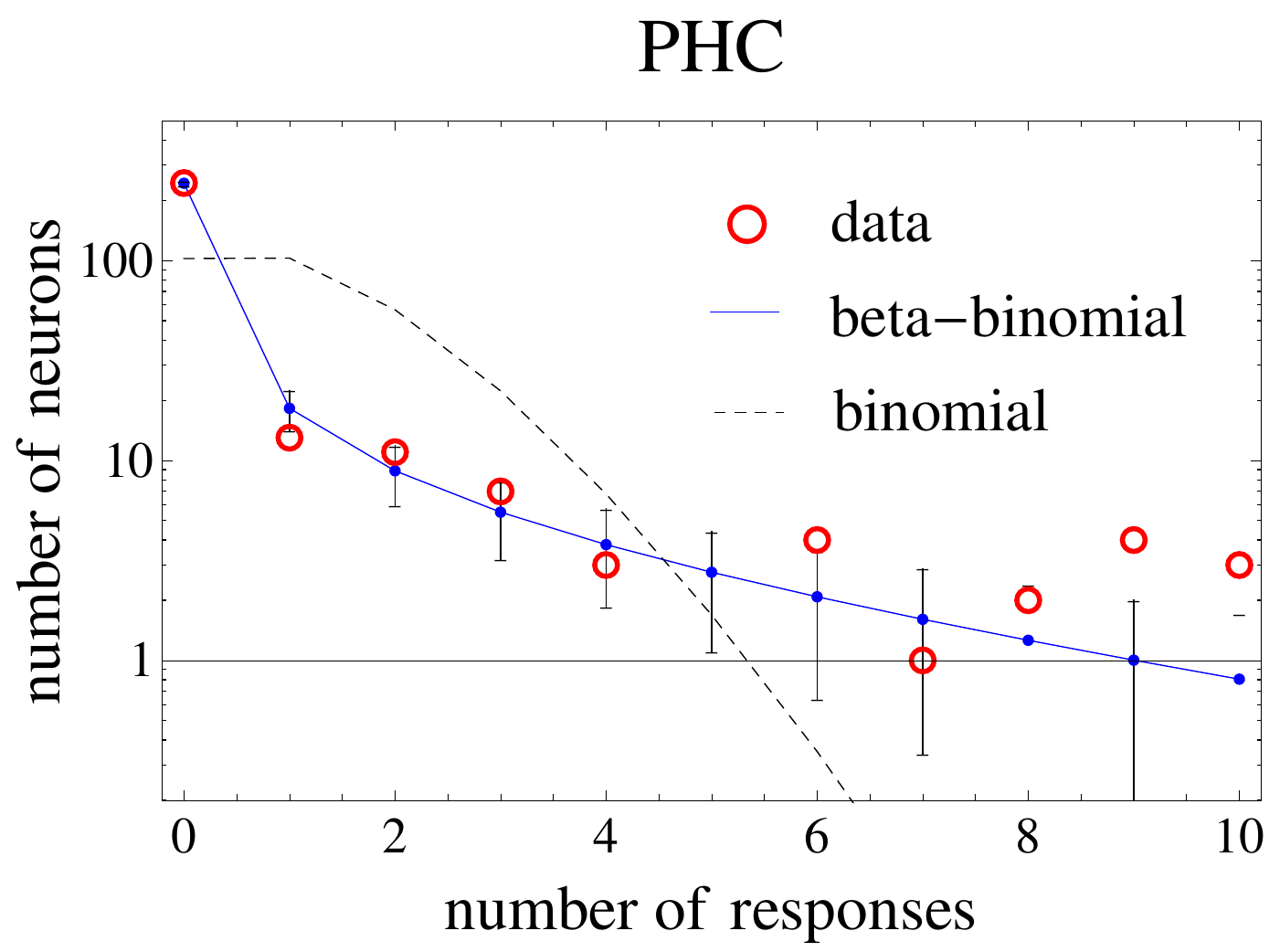} 
\end{tabular}
\caption{Best fits to data recorded in four regions of the human MTL assuming all recorded units consist of a single neuron. The red circles indicate the experimental values from Table \ref{table:rawdata} and
  blue dots connected by lines indicate the best-fit predictions for the
  expectation values of $n_k$ with double-neuron units included predicted by the beta-binomial model. The dotted lines indicate best fits for a pure binomial model i.e. assuming all cells have the same sparsity. Note the log scale on the y-axis.}
\label{fig:MultiUnitplots}
\end{figure*}

In this section, we use the beta model to analyze data taken from the human MTL presented in \cite{mormann}. The data is recorded from 1194 Hipp units, 844 EC units, 947 Amy units, and 293 PHC units. The patients were shown on average 97 images of familiar stimuli, presented in random order. The $n_k$ for each region are given in Table \ref{table:rawdata}.
\begin{table*}
  \centering
  \begin{tabular}{c|c|c|c|c|c|c|c|c|c|c|c|c|c|c|c}
   	& $n_0$ & $n_1$ & $n_2$ & $n_3$ & $n_4$ & $n_5$ & $n_6$ & $n_7$ & $n_8$ & $n_9$ & $n_{10}$ & $n_{11}$ & $n_{12}$ & $n_{13}$ & $n_{14}$ \\ \hline
    Hipp & $1019$ & $113$ & $30$ & $17$ & $7$ & $4$ & $1$ & $2$ & $0$ & $0$ & $0$ & $0$ & $1$ & $0$ & $0$ \\
    EC & $761$ & $45$ & $15$ & $9$ & $4$ & $8$ & $0$ & $0$ & $1$ & $0$ & $0$ & $0$ & $0$ & $1$ & $0$ \\
    Amy & $842$ & $61$ & $17$ & $15$ & $3$ & $3$ & $1$ & $0$ & $1$ & $1$ & $0$ & $0$ & $0$ & $1$ & $2$ \\
    PHC & $244$ & $13$ & $11$ & $7$ & $3$ & $0$ & $4$ & $1$ & $2$ & $4$ & $3$ & $0$ & $1$ & $0$ & $0$ \\ 
  \end{tabular}
  \caption{Number of units $n_k$ responding to $k$ images as
    reported by \cite{mormann} in four MTL regions.} 
  \label{table:rawdata}
\end{table*}
Maximizing (\ref{likelihood}) with the respect to the $a$ and $b$ yields the parameter values given in Table \ref{MLEvalues} for the four regions. The goodness of each fit was assessed using $\chi^2$ test. Plots of the fits against the data are given in figure \ref{fig:MultiUnitplots}. 

\begin{figure*}
\centering
\begin{tabular}{c@{\hspace*{6mm}}c}
    \includegraphics[scale=0.4]{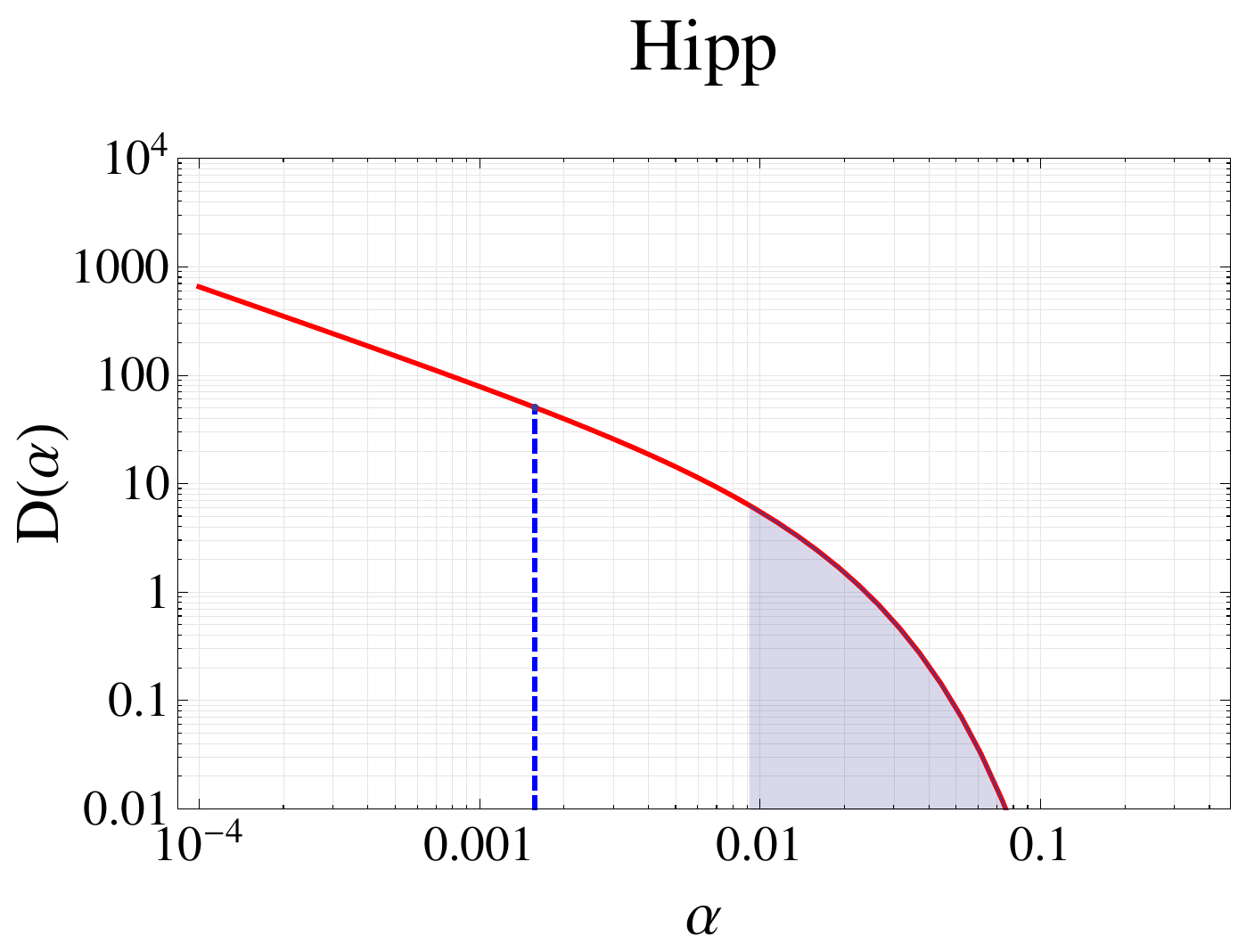} &  \includegraphics[scale=0.4]{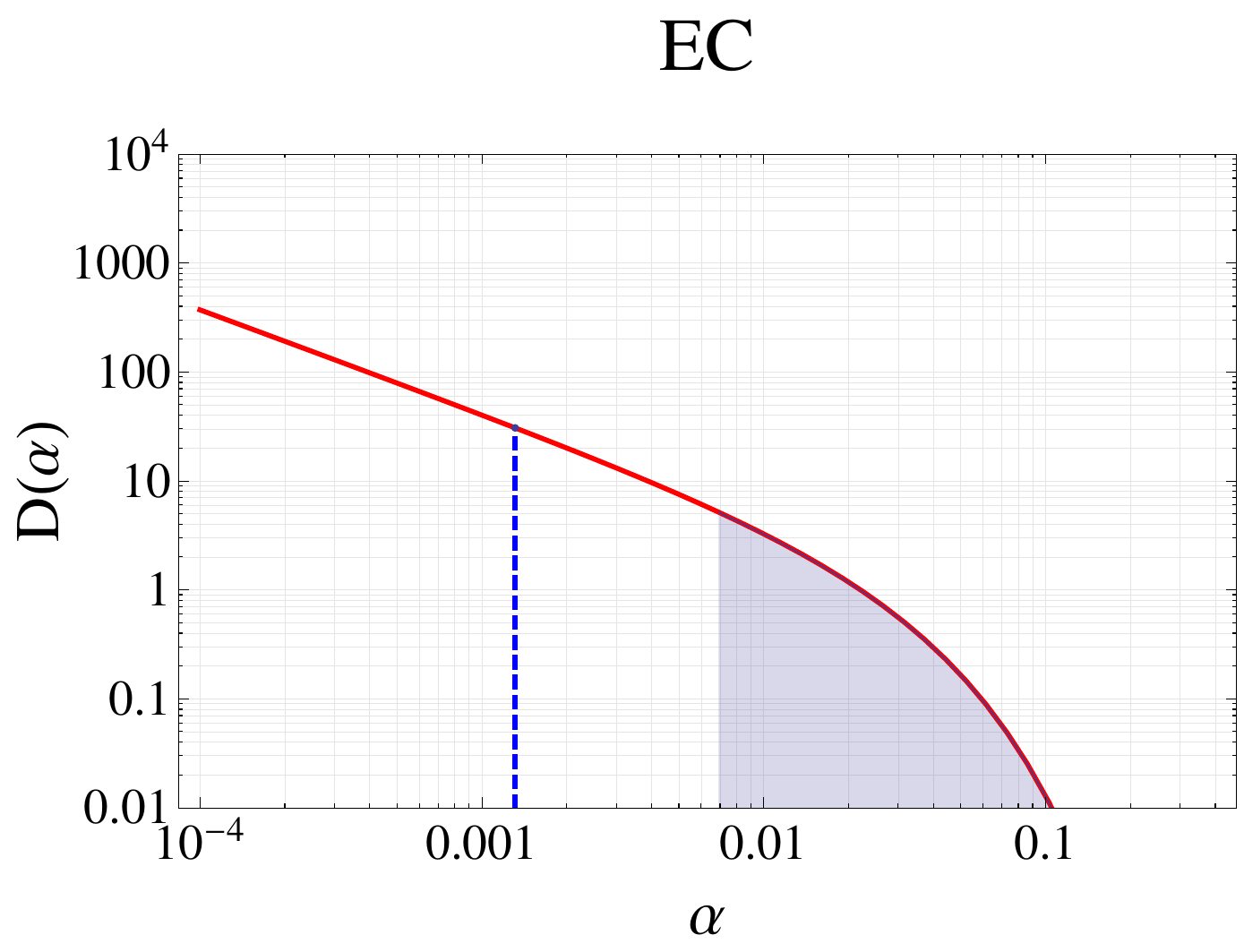} 
    \\ \includegraphics[scale=0.4]{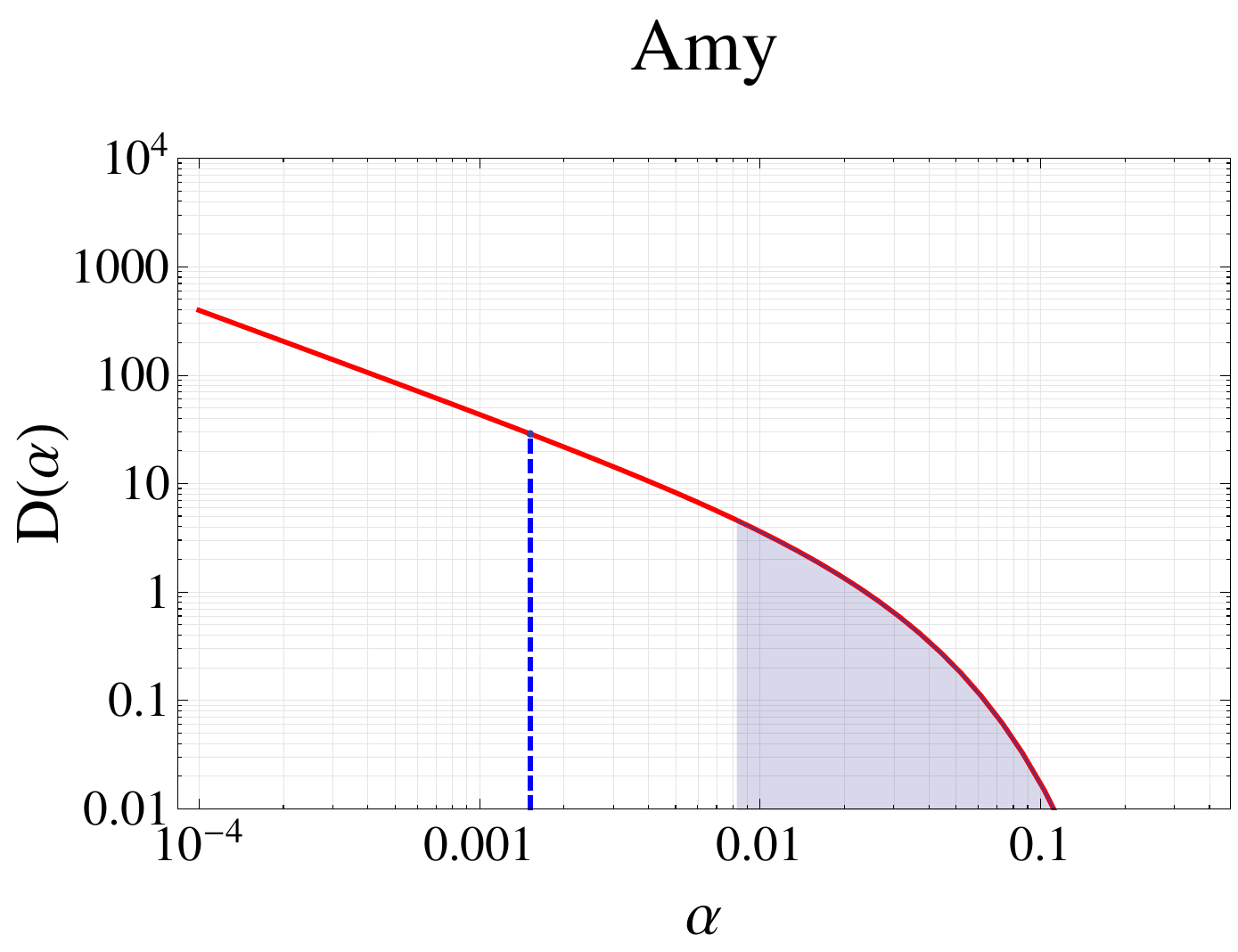} &  \includegraphics[scale=0.4]{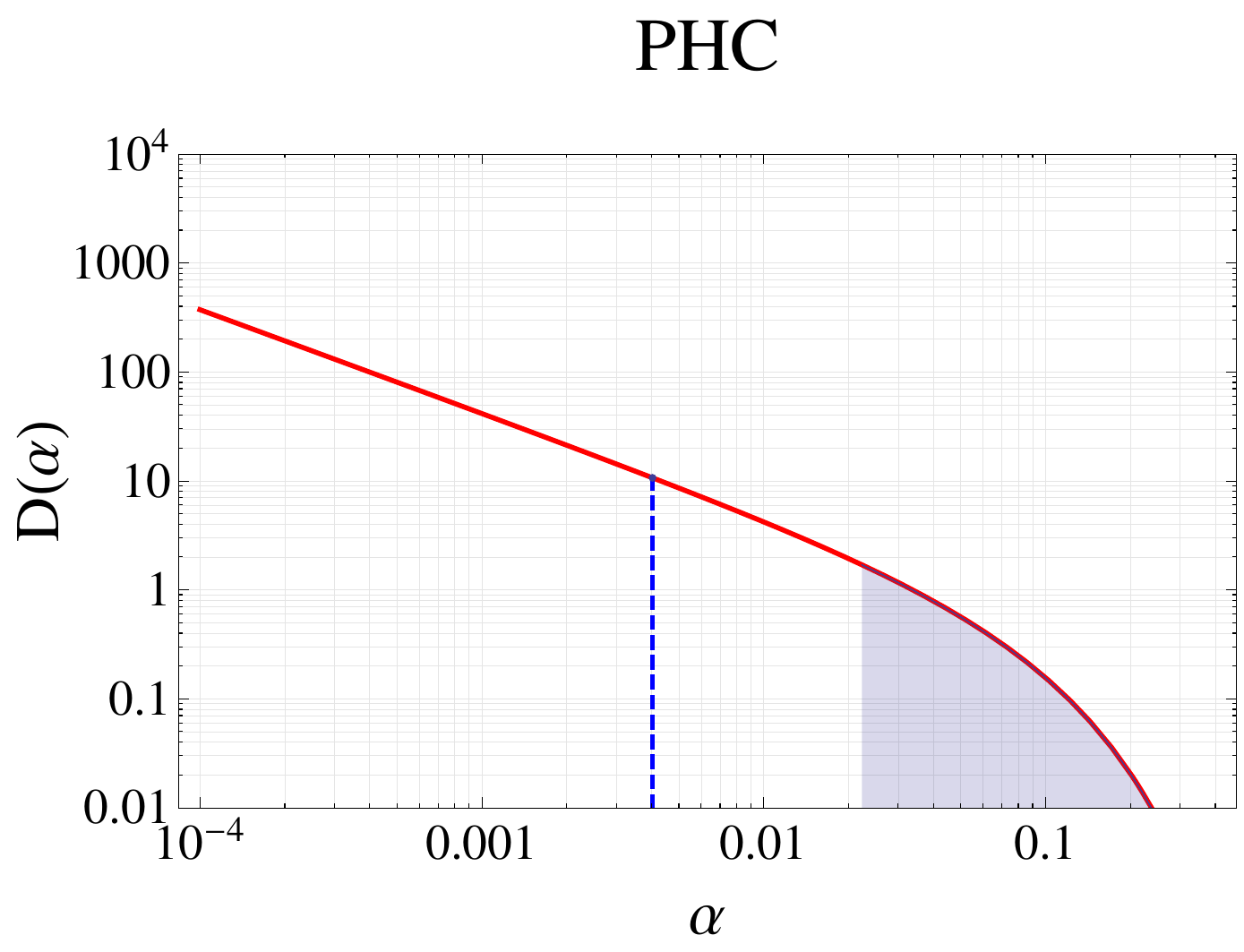} 
\end{tabular}
\caption{Sparsity distributions across neurons in each region predicted by the multi-unit model consisting of single units and double units. The plots are shown using log-log axes to indicate power law behavior at low sparsity. The shaded region indicates the upper 5\% tail of the sparsity distribution, i.e. the model predicts 95\% of the neurons in each region have a sparsity left of the shaded area. The verticle dotted lines indicate the mean sparsities in each region.}
\label{fig:dist}
\end{figure*}

\begin{table*}
\centering
\begin{tabular}{cc}
\textit{Single-Unit Model} & \textit{Multi-Unit Model} \\ \\
%\begin{tabular}{c|c|c|c|c}
%	 & PHC & EC & Hipp & Amy \\ \hline
%	$a$ & 0.08  & 0.08   &   0.17   &  0.09   \\ \hline
%	$b$ &  12 &  36  &   66   &  34   \\ \hline
%	$\chi^2(10)$ & 13.7  &  12.8  & 4.38     & 8.52 
%\end{tabular}
\begin{tabular}{c|c|c|c|c}
	 & Hipp & EC & Amy & PHC \\ \hline
	$a$ & 0.17  & 0.08   &   0.09   &  0.08   \\ \hline
	$b$ &  66 &  36  &   34   &  12   \\ \hline
	$\chi^2(5)$ & 6.1  &  1.9  & 8.6     & 3.3 
\end{tabular}
\hspace{5pt} & \hspace{5pt}
%\begin{tabular}{c|c|c|c|c}
%	 & PHC & EC & Hipp & Amy \\ \hline
%	$a$ & 0.05  & 0.05   &   0.11   &  0.05   \\ \hline
%	$b$ &  13 &  36  &   67   &  34   \\ \hline
%	$\chi^2(10)$ & 17.0  &  12.8  & 4.4     & 8.0 
%\end{tabular} 
\begin{tabular}{c|c|c|c|c}
	 & Hipp & EC & Amy & PHC \\ \hline
	$a$ & 0.11  & 0.05   &   0.05   &  0.05   \\ \hline
	$b$ &  67 &  36  &   34   &  13   \\ \hline
	$\chi^2(5)$ & 2.1  &  0.56  & 5.2     &  2.7 
\end{tabular} 
\end{tabular}
\caption{Best fit parameters of the beta distribution $a$ and $b$ for the four MTL regions. Left-hand table gives values assuming all recorded units consist of a single neuron. Right-hand table gives values assuming a mixture of single and double neuron units. $\chi^2$ values evaluated for $k=1$ to $k=10$.}
\label{MLEvalues}
\end{table*}

All four regions are successfully fit by both the beta model containing single units and the improved model containing a mixture of single and double neuron units ($\chi^2$ values given in table \ref{MLEvalues}). This suggests that the data is consistent with the notion that the sparsity of human MTL neurons follows a beta distribution, given in equation \ref{beta}, with $a<1$ and $b>1$. The sparsity distribution is far from uniform, consistent with the result in \cite{twopop} when the two-population model was used. 

Including the double units in the model had little effect on the goodness of fit, though the value of the $a$ parameter is brought closer to zero. This results in a lower mean sparsity compared with assuming only single neuron units.

Plots of the best-fit sparsity distributions from the mixture model are given in figure \ref{fig:dist}. The means of the distribution in each of the regions are: $\left\langle \alpha \right\rangle=1.6\times10^{-3}$ in Hipp; $1.5\times10^{-3}$ in EC; $1.5\times10^{-3}$ in Amy; $4.0\times10^{-3}$ in PHC. In each region, $a$ is close to zero, producing a near-divergence in the neuronal sparsity distribution as $\alpha\rightarrow0$. This indicates a large population of extremely sparse neurons, consistent with the results of \cite{twopop}. For these fits, roughly 95\% of the MTL neurons falls within the power-law regime, as is shown by the linear region of the log-log plots in figure \ref{fig:dist}.

The sparsity distribution in the PHC differed quantitatively from the fits in the other three regions. Firstly, the model predicts that neurons of the PHC has the highest mean sparsity, $\left\langle \alpha \right\rangle=4\times10^{-3}$, compared to the other three regions which have mean sparsities clustered near $1.5\times10^{-3}$. Also, the tail of the sparsity distribution extends to larger $k$ for the PHC than it does for the other three regions, as shown in figure \ref{fig:dist}. This is consistent with the observation that the selectivities of neurons tend to increase as information proceeds down the ventral visual pathway from PHC to the EC, Hipp and Amy \cite{mormann}. 

%=======================

\subsection{A Model for Representational Learning by Preferential Attachment}

One advantage of using statistical inference to estimate the underlying distribution of sparsity, $D\left(\alpha\right)$, is that the form of the distribution can suggest a particular generating mechanism. Beta distributions with near power law behavior arise as limiting distributions in numerous ``rich-get-richer'' schemes. For example, both preferential attachment processes on growing networks, where nodes with a high degree are more likely to receive edges from newly added nodes \cite{barabasi, newman}, and Polya urn schemes \cite{mahmoud}, where the proportion of balls of a particular color grows whenever a ball of that color is sampled from the urn, both yield beta distributions as $t \rightarrow \infty$. 

In this section, we consider the bi-layered network model studied in Peruani et al. \cite{peruani}. In their model, the bottom layer of the network contains $\mathcal{N}$ nodes which remain fixed in number, while the top layer consists of $t$ nodes that are added one at a time, starting from $t=0$. As nodes are added, they connect with nodes in the bottom layer with a higher probability of attaching to nodes with a large degree.  

We interpret the fixed bottom layer as the binary neurons of the MTL responsible for object encoding, and we interpret the top layer as stimuli related to familiar concepts which have been previously encoded into memory. The addition of a node in the top layer indicates a new concept that is to be coded into memory, e.g. an unfamiliar person to whom you have just been introduced. An edge between stimulus $i$ in the top layer and neuron $j$ in the bottom layer indicates that neuron $j$ is part of the code for $i$ (i.e. neuron $j$ activates whenever stimulus $i$ is presented to the organism). The attachment procedure for new nodes represents the complex neurological learning processes by which new concepts are coded onto the neural substrate of the human MTL. Thus, the growth of the top layer with the attachment of edges to the bottom layer is a model for representational learning of a binary code. The sparsity of neuron $j$, $\alpha_j$, is given by
\begin{equation}
\alpha_j=\frac{k_j^t}{t}
\label{sparsity:pref}
\end{equation}
where $k_j^t$ is the node degree of neuron $j$ after $t$ stimuli have been added in the top layer.

Following Peruani, et al. \cite{peruani}, the network is grown by adding a node to the top layer at each time step, $t$, and then attaching it to $\mu$ nodes in the bottom layer. In our model, $\mu$ is the code word length of each stimulus and is assumed to be fixed. We assume $\mu<<\mathcal{N}$. As each of the $\mu$ edges is added, the probability $A\left(k_j^t\right)$ that it attaches to neuron $j$ with node degree $k_j^t$ is defined by
\begin{equation}
A\left(k_j^t\right) = \frac{1}{C\left(t\right)} \left( \gamma k_j^t + 1 \right) 
\label{attachment}
\end{equation}
where $\gamma$ is a parameter that determines the influence of node degree on the attachment process and $C\left(t\right)$ is a normalization constant. For $\gamma=0$, the edges are attached at random to the neurons with $A\left(k_j^t\right)=\frac{1}{\mathcal{N}}$ for each $j$. In the case where $\mu<<\mathcal{N}$, the probability that neuron $j$ connects to the stimulus added at time $t$ after all $\mu$ edges have been connected is approximately
\begin{equation}
A_{\mu}\left(k_j^t\right) = \frac{1}{C\left(t\right)} \mu \left( \gamma k_j^t + 1 \right) 
\label{approx.kernel}
\end{equation}
When $\gamma=0$, a newly added stimulus attaches to neuron $j$ with probability $A_{\mu}\left(k_j^t\right) = \frac{\mu}{\mathcal{N}}$. 

The network is grown starting from $t=0$, with no stimuli in the top layer and all node degrees in the bottom layer equal zero. Equation (\ref{approx.kernel}) defines the attachment process as each stimulus is added. We are interested in the sparsity distribution, $D\left( \frac{k}{t} \right)$ of neurons in the bottom layer after a large number of stimuli have been learned. Peruani et al. \cite{peruani} showed that for $\gamma>0$, in the limit of large $t$,
\begin{equation}
D\left( \frac{k}{t} \right)=\frac{1}{B\left( r, s \right)} \left(\frac{k}{t}\right)^{r-1} \left(1-\frac{k}{t}\right)^{s-1}
\label{beta.dist.pref}
\end{equation}
where $r=\frac{1}{\gamma}$ and $s=\frac{\mathcal{N}}{\gamma \mu} - \frac{1}{\gamma}$. 

In the case of purely random attachment ( $\gamma=0$), the distribution $D\left( \frac{k}{t} \right) \rightarrow \delta \left(\frac{k}{t} - \frac{\mu}{\mathcal{N}} \right)$ as $t \rightarrow \infty$, i.e. the neuronal sparsity is the same for all neurons. Purely random attachment is inconsistent with the data fit above, as can be seen from the dotted lines in \ref{fig:MultiUnitplots}.

If we treat the data analyzed above as a random sample of $N$ neurons from the bottom layer and $S$ stimuli from the top layer, then we can match the beta parameters $r$ and $s$ with the fits reported above. This gives an estimate of roughly $10 \leq \gamma \leq 20 $ and $\frac{1}{600} \leq \frac{\mu}{\mathcal{N}} \leq \frac{1}{250}$. These values for $\gamma$ suggest from equation \ref{approx.kernel} that preferential attachment plays a large role in assigning new stimuli to MTL neurons. In other words, it is indirect evidence that neurons are not randomly assigned to new concepts, but rather new concepts are more likely to be coded onto neurons that have previously been assigned to previously learned concepts.

%=======================

\section{Discussion}
\label{sec:discussion}

\subsection{Relation to Previous Work}

This paper builds upon previous work \cite{collins,twopop} within which neurons were assumed to be split into two discrete populations: a sparse population comprising roughly $5\%-10\%$ of the MTL neurons, with a sparsity on the order of $10^{-2}$; and an ultra-sparse population comprising the remaining $90\%-95\%$, with a sparsity on the order of $10^{-3}$. There, the neurons in each population were assumed to have the same sparsity value. This two-population model produced good fits in the Hipp and the EC, but produced poorer fits in the Amy and the PHC. The beta model developed in this paper is a continuous distribution, and it fits all four regions adequately, including the Hipp and the EC.

Thus, there are two radically different sparsity distributions that are consistent with the single-unit responses from the Hipp and EC. In order to produce good fits to the data, which shows very large $n_0$ and $n_1$ bins compared with the bins at higher $k$, the sparsity distributions at small $\alpha$ are largely determined by these two bins. There is insufficient statistical power to distinguish between different sparsity distributions in this low sparsity regime. 

To illustrate the connection between the beta model and the two-pop model, the shaded region seen in figure \ref{fig:dist}, roughly corresponds to the sparse population (population labeled ``D'' in \cite{twopop}), while the neurons in the unshaded region are analogous to the ultra-sparse population (labeled ``US'' in \cite{twopop}). In the two-population model, the distribution would have a Dirac-delta function located in the shaded region and another located in the unshaded region, representing the two populations. 

Consequently, the exact form of the beta distribution should not be taken too literally, as there are likely to be other continuous distributions that are consistent with the data. However, one would still expect similar overall behaviors regardless of which distribution is chosen. The advantage of the beta distribution is that it yields compact analytical results in the likelihood analysis for both the single unit model and the multi-unit model. 

\subsection{Similarity to Place Cell Statistics}

Recent experiments have revealed that place cells in the rat hippocampus display skewed activity, in which the number of place fields recruited by place cells exhibits a heavier tail than what would be predicted if the cells all recruited place fields at the same rate \cite{rich}. In the experiment performed by Rich et al. \cite{rich}, the recruitment of place fields among the cells of CA1 obeys a skewed gamma-poisson process, in which each cell acquires place fields according to a poisson process, but the rate parameter for each cell is sampled from a gamma distribution. This results in a poisson distribution of place fields per cell mixed by a gamma distribution. In this paper, the number of concpet fields of MTL neurons are shown above to be be distributed by a binomial distribution mixed with a beta distribution. The gamma-poisson distribution and beta-binomial distributions are closely related, with the gamma-poisson being a limiting case of the beta-binomial. 

Place cells of the rat and concept cells in humans share many characteristics \cite{qq.concept}, and observing nearly identicle distributions of activity in both populations suggests that the place cell code and the concept cell code likely arise from similar underlying processes. In other words, the results suggest that the recruitment of place fields by the place cells of the rat hippocampus and the recruitment of ``concept fields'' by the concept cells of the human MTL are two manifestations of the same neural mechanism, despite coding for spatial information vs conceptual information. 

Furthermore, finding nearly identicle distributions in different species across cells with strikingly different receptive fields lends more credence to the idea that skewed distributions in general play a fundamental role in neural functioning \cite{buzsaki.2015, buzsaki.2014}.

One possible mechanism for generating skewed distributions, especially distributions that have that have approximate power-law behavior, are the various cumulative advantage or ``rich get richer'' schemes such as preferential attachment processes on growing networks explored above. Another possibility, briefly explored in \cite{rich} is that the non-uniform recruitment of receptive fields arises from intrinsic cell differences, such as non-uniform excitability and pre-synaptic inputs. I suggest that these two hypotheses are not mutually exclusive, and that both may play a role in generating the observed sparse, skewed distributions. %\cite{barabasi, newman}, and Polya urn processes \cite{mahmoud}. In particular, Peruani et al. \cite{peruani} have shown that the distribution of node degrees on a two-layered network grown by preferential attachment processes approaches a beta distribution as one layer of the network grows and the other is held fixed. We apply their result to our fits, treating the fixed layer and the growing layer of their network as the neurons of the human MTL, and the stimuli to be represented, respectively. As a result, we suggest preferential attachment as a possible mechanism for learning sparse representations of sensory stimuli.

\subsection{Impact of Silent Cells}

The spike sorting techniques for single unit-recordings only detect a small fraction of the neurons within range of the electrode. The remaining neurons, constituting perhaps as much as $90\%$ of the overall cells, remain completely silent during the experiment and thus are missed completely by the spike sorting algorithm \cite{waydo, henze, thompson, olshausen1}. This means that the sample of neurons is biased in favor of more active cells. 

To clarify, these silent cells, or ``neural dark matter,'' are distinct from the $n_0$ cells reported in \cite{mormann} which did not produce above threshold firing rates for any of the presented stimuli. The population of $n_0$ cells still emitted spikes and thus were detected by the spike sorting techniques used. The silent cells on the other hand, emitted no spikes, and thus could not be detected. In effect, these cells should be included in the $n_0$ bin to give a more biased estimate of the sparsity distribution. The result on the fits would be to lower the value of $a$, bringing it even closer to zero.

Some studies estimate the silent cell population to be as high as a factor of ten \cite{waydo}. That is, there are perhaps ten silent cells for each recorded cell. Using this factor of cells added to the $n_0$ bin, maximum likelihood analysis yields $a=0.007$ and $b=55$ in the hippocampus, with $\chi^2=1.2$. This brings the sparsity distribution much closer to a power law, indicating a considerably sparser code.

\begin{acknowledgements}
I would like to thank John Collins and Mike Skocik for valuable commentary and discussions regarding the content of this manuscript.
\end{acknowledgements}

%=============================================
%=============================================

\appendix
\section{Derivation of multi-unit response probability}
In this section we derive equation (\ref{multibbn}) starting from the integral given in equation (\ref{multiepsilon}):
\begin{align}
\epsilon_{(2),k}=\int_0^1{d\alpha_1}\int_0^1{ d\alpha_2} &  \big[ D_{a,b}\left( \alpha_1 \right) D_{a,b}\left( \alpha_2 \right)\, \times \nonumber\\
& \binom{S}{k} \, \alpha_{unit}^{k} (1-\alpha_{unit}) ^{S-k} \big]
\label{app}
\end{align}
where $D_{a,b}\left( \alpha \right)$ is the beta distribution PDF given in equation (\ref{beta}), and $\alpha_{unit}$ is the effective sparsity for the double unit given by equation (\ref{unitalpha}).

To evaluate the integral, we first note that:
\begin{equation} 
\left( 1-\alpha_{unit} \right)^{S-k}  = [ \left( 1-\alpha_1 \right) \left( 1-\alpha_2 \right) ]^{S-k}
\end{equation} 
Then, we expand the factor $\alpha_{unit}^k$ in (\ref{app}) using the binomial theorem:
\begin{align}
\alpha_{unit}^k & = \left( \alpha_1 + \alpha_2 \left( 1-\alpha_1 \right) \right)^k \nonumber \\
& =\sum_{j=0}^k{\binom{k}{j} \alpha_1^j [\alpha_2 \left(1-\alpha_1 \right)]^{k-j}} 
\label{expansion}
\end{align}
Using these results and equation (\ref{beta}), we can write equation (\ref{app}) as a sum of separable integrals
\begin{align}
\epsilon_{(2),k}=& \frac{1}{[B\left(a,b\right)]^2} \, \binom{S}{k} \, \sum_{j=0}^k  { \big\{ \, \binom{k}{j} } \,  \nonumber\\
& \times \int_0^1{ d\alpha_1 \, \alpha_1^{a-1+j} \, \left(1-\alpha_1 \right)^{b-1+S-j}}  \nonumber\\
&\times \int_0^1{d\alpha_2  \, \alpha_2^{a-1+k-j} \, \left(1-\alpha_2 \right)^{b-1+S-k}}\big\} 
\end{align}
The integrals can now be evaluated as beta functions
\begin{align}
\epsilon_{(2),k}= \frac{1}{[B\left(a,b\right)]^2} \,& \binom{S}{k} \, \sum_{j=0}^k  { \big\{ \, \binom{k}{j} } \,  \nonumber\\
& \times B\left(a+j , b+S-j \right) \nonumber \\
& \times B\left( a+k-j , b+S-k \right) \big\}
\end{align}
yielding the result given by equation (\ref{multibbn}).
%=============================================

\section{Gamma-poisson distribution as a limiting case of the beta-binomial distribution}

\label{appendix.gamma-poisson}

In this appendix, I show that the gamma-poisson distribution is a limiting case of the beta-binomial distribution. I show it here because a convenient reference showing this result could not be located.

The beta-binomial distribution is a mixture distribution of a binomial distribution, with parameters $S$ and $0<\alpha<1$, where the probability parameter $\alpha$ is sampled from a beta distribution with parameters $a>0$ and $b>0$. The PMF for the binomial distribution and the PDF for the beta distributions are given respectively:
\begin{equation}
P\left(K=k\right)= \binom{S}{k} \alpha^k  \left(1-\alpha\right)^{S-k}
\end{equation}
and
\begin{equation}
D\left(\alpha\right)= \frac{\alpha^{a-1}  \left(1-\alpha\right)^{b-1}}{B\left(a,b\right)}
\label{b}
\end{equation}
respectively.

Thus, the PMF of the beta-binomial distribution with parameters $S$, $a$, and $b$ is given by:
\begin{align}
		P\left(K=k\right)&=\frac{\binom{S}{k}}{B\left(a,b\right)}  \int_0^1{\text{d}\alpha  \: \alpha^k  \left(1-\alpha\right)^{S-k}   \alpha^{a-1}  \left(1-\alpha\right)^{b-1}} \nonumber \\
		&=\binom{S}{k} \frac{B\left(a+k,b+S-k\right)}{B\left(a,b\right)}.
\label{bb}
\end{align}
The gamma-poisson distribution, more commonly called the negative binomial distribution, is a Poisson distribution with parameter $\lambda>0$ where $\lambda$ is sampled from a gamma distribution with parameters $r>0$ and $\beta>0$. The PMF of the Poisson distribution and the PDF of the gamma distribution is given:
\begin{equation}
P\left(K=k\right)= \frac{\lambda^{k}e^{-\lambda}}{k!}
\end{equation}
and 
\begin{equation}
g\left(\lambda\right) = \frac{r^\beta}{\Gamma(\beta)} \lambda^{\beta-1}e^{-r \lambda}.
\end{equation}
and the gamma-poisson distribution with parameters $r$ and $\beta$ is then given by
\begin{align}
P\left(K=k\right) &= \frac{1}{k!}\frac{r^{\beta}}{\Gamma(\beta)} \int_0^{\infty}{\text{d}\lambda  \: \left[ \lambda^k e^{-\lambda} \right] \left[ \lambda^{\beta-1} e^{-r \lambda} \right] } \nonumber \\
&= \frac{1}{k!}\frac{r^{\beta}}{\Gamma(\beta)} \int_0^{\infty}{\text{d}\lambda  \:  \lambda^{k+\beta-1} e^{-\lambda(1+r)} } \nonumber \\
&= \frac{1}{k!} \frac{r^{\beta}}{\Gamma(\beta)} \int_0^{\infty}{\frac{\text{d}x}{r+1} \left(\frac{x}{r+1}\right)^{k+\beta-1} e^{-x}} \nonumber \\
&= \frac{1}{k!} \frac{r^{\beta}}{\Gamma(\beta)} \frac{\Gamma(k+\beta)}{\left(r+1\right)^{k+\beta}}
\label{gp}
\end{align}
To show that \ref{bb} yields \ref{gp} as a limiting case, we begin by observing that the Poisson distribution is a limiting case of the binomial distribution as $S\rightarrow \infty$ while holding the mean, $\langle k \rangle = S\alpha$ constant. The Poisson parameter $\lambda$ is the expected response, i.e. $\lambda=\langle k\rangle$.

Similarly, the gamma-Poisson distribution is the limiting case of the beta-binomial distribution as $S \rightarrow \infty$ while ensuring the mean $\langle k \rangle = S \langle \alpha \rangle$ for the beta-binomial distribution is finite. From equation (\ref{b}), we see that 
\begin{align}
\langle \alpha \rangle &= \frac{1}{B(a,b)} \int_0^1{\alpha^a \left(1-\alpha\right)^{b-1}} \nonumber \\
&= \frac{B(a+1,b)}{B(a,b)} \nonumber \\
&= \frac{a}{a+b}
\end{align}
 So, 
\begin{equation}
\langle k \rangle = S\langle \alpha \rangle =\frac{S a}{a+b}.
\end{equation} 
One way to ensure $\langle k \rangle$ is finite as $S \rightarrow \infty$, is for the parameter $b$ to approach infinity as a linear function of $S$ by setting $b=r S$. The constant $r>0$ is arbitrary. It will be shown that that $r$ matches the parameter of the gamma distribution above when the limit is taken.

Before taking the limit by setting $b=r S$ in equation (\ref{bb}),  it is useful to write the beta functions and binomial coefficient in terms of gamma functions using the identities
\begin{equation}
\binom{S}{k}=\frac{\Gamma(S+1)}{\Gamma(k+1) \Gamma(S-k+1)}
\label{bin.gamma}
\end{equation}
and
\begin{equation}
B\left(x,y\right)= \frac{\Gamma(x) \Gamma(y)}{\Gamma(x+y)}
\end{equation}
Using these identities, we get write equation (\ref{bb}) as
\begin{align}
P\left(K=k\right) &= \frac{\Gamma(S+1)}{\Gamma(k+1) \Gamma(S-k+1)}  \nonumber \\ 
& \times \frac{\Gamma(a+k) \Gamma(S-k+b)}{\Gamma(a+b+S)} 
\times \frac{\Gamma(a+b)}{ \Gamma(a) \Gamma(b)} \nonumber \\ \nonumber \\
&= \frac{\Gamma(S+1)}{\Gamma(k+1) \Gamma(S-k+1)} \nonumber \\
&\times \frac{\Gamma(a+k) \Gamma(S(1+r)-k)}{\Gamma(a+S(1+r))} \times \frac{\Gamma(a+rS)}{ \Gamma(a) \Gamma(rS)}
\end{align}
Now, to take the limit as $S \rightarrow \infty$, we use Stirling's approximation applied to ratios of gamma functions:
\begin{equation}
\frac{\Gamma(x+\gamma)}{\Gamma(x+\omega)} \approx x^{\gamma-\omega}
\label{stirling}
\end{equation}
for large $x$. We get:
\begin{align}
P\left(K=k\right) &= \frac{S^k}{\Gamma(k+1)} \times \frac{\Gamma(a+k)}{\left[S(1+r)\right]^{a+k}} 
\times \frac{(rS)^{a}}{\Gamma(a)} \nonumber \\
&= \frac{1}{k!} \times \frac{\Gamma(a+k)}{(1+r)^{a+k}} \times \frac{r^a}{\Gamma(a)}
\label{gp.fin}
\end{align}
Comparing equations (\ref{gp.fin}) and (\ref{gp}), we see that they match, and that $a=\beta$. This concludes the derivation.
%=============================================

\bibliography{betarefs}

\begin{thebibliography}{38}
\expandafter\ifx\csname natexlab\endcsname\relax\def\natexlab#1{#1}\fi
\expandafter\ifx\csname bibnamefont\endcsname\relax
  \def\bibnamefont#1{#1}\fi
\expandafter\ifx\csname bibfnamefont\endcsname\relax
  \def\bibfnamefont#1{#1}\fi
\expandafter\ifx\csname citenamefont\endcsname\relax
  \def\citenamefont#1{#1}\fi
\expandafter\ifx\csname url\endcsname\relax
  \def\url#1{\texttt{#1}}\fi
\expandafter\ifx\csname urlprefix\endcsname\relax\def\urlprefix{URL }\fi
\providecommand{\bibinfo}[2]{#2}
\providecommand{\eprint}[2][]{\url{#2}}

\bibitem[{\citenamefont{Barlow}(1972)}]{barlow}
\bibinfo{author}{\bibfnamefont{H.~B.} \bibnamefont{Barlow}},
  \bibinfo{journal}{Perception} pp. \bibinfo{pages}{795--8}
  (\bibinfo{year}{1972}).

\bibitem[{\citenamefont{F\"oldi\'ak and Endres}(2008)}]{foldiak1}
\bibinfo{author}{\bibfnamefont{P.}~\bibnamefont{F\"oldi\'ak}} \bibnamefont{and}
  \bibinfo{author}{\bibfnamefont{D.}~\bibnamefont{Endres}},
  \bibinfo{journal}{Scholarpedia} \textbf{\bibinfo{volume}{3(1)}},
  \bibinfo{pages}{2984} (\bibinfo{year}{2008}).

\bibitem[{\citenamefont{F\"oldi\'ak and Young}(2002)}]{foldiak2}
\bibinfo{author}{\bibfnamefont{P.}~\bibnamefont{F\"oldi\'ak}} \bibnamefont{and}
  \bibinfo{author}{\bibfnamefont{M.}~\bibnamefont{Young}}, in
  \emph{\bibinfo{booktitle}{Handbook of brain theory and neural networks}},
  edited by \bibinfo{editor}{\bibfnamefont{M.}~\bibnamefont{Arbib}}
  (\bibinfo{publisher}{MIT Press}, \bibinfo{address}{Cambridge, MA},
  \bibinfo{year}{2002}), pp. \bibinfo{pages}{1064--1068}.

\bibitem[{\citenamefont{Bowers}(2009)}]{bowers}
\bibinfo{author}{\bibfnamefont{J.}~\bibnamefont{Bowers}},
  \bibinfo{journal}{Psychol. Rev.} \textbf{\bibinfo{volume}{116}},
  \bibinfo{pages}{220} (\bibinfo{year}{2009}).

\bibitem[{\citenamefont{Olshausen and Field}(2004)}]{olshausen.sparse}
\bibinfo{author}{\bibfnamefont{B.~A.} \bibnamefont{Olshausen}}
  \bibnamefont{and} \bibinfo{author}{\bibfnamefont{D.~J.} \bibnamefont{Field}},
  \bibinfo{journal}{Current opinion in neurobiology}
  \textbf{\bibinfo{volume}{14}}, \bibinfo{pages}{481} (\bibinfo{year}{2004}).

\bibitem[{\citenamefont{Willmore and Tolhurst}(2001)}]{willmore2001}
\bibinfo{author}{\bibfnamefont{B.}~\bibnamefont{Willmore}} \bibnamefont{and}
  \bibinfo{author}{\bibfnamefont{D.~J.} \bibnamefont{Tolhurst}},
  \bibinfo{journal}{Network: Computation in Neural Systems}
  \textbf{\bibinfo{volume}{12}}, \bibinfo{pages}{255} (\bibinfo{year}{2001}).

\bibitem[{\citenamefont{Willmore et~al.}(2011)\citenamefont{Willmore, Mazer,
  and Gallant}}]{willmore2011}
\bibinfo{author}{\bibfnamefont{B.~D.} \bibnamefont{Willmore}},
  \bibinfo{author}{\bibfnamefont{J.~A.} \bibnamefont{Mazer}}, \bibnamefont{and}
  \bibinfo{author}{\bibfnamefont{J.~L.} \bibnamefont{Gallant}},
  \bibinfo{journal}{Journal of neurophysiology} \textbf{\bibinfo{volume}{105}},
  \bibinfo{pages}{2907} (\bibinfo{year}{2011}).

\bibitem[{\citenamefont{Perez-Orive et~al.}(2002)\citenamefont{Perez-Orive,
  Mazor, Turner, Cassenaer, Wilson, and Laurent}}]{perez}
\bibinfo{author}{\bibfnamefont{J.}~\bibnamefont{Perez-Orive}},
  \bibinfo{author}{\bibfnamefont{O.}~\bibnamefont{Mazor}},
  \bibinfo{author}{\bibfnamefont{G.~C.} \bibnamefont{Turner}},
  \bibinfo{author}{\bibfnamefont{S.}~\bibnamefont{Cassenaer}},
  \bibinfo{author}{\bibfnamefont{R.~I.} \bibnamefont{Wilson}},
  \bibnamefont{and} \bibinfo{author}{\bibfnamefont{G.}~\bibnamefont{Laurent}},
  \bibinfo{journal}{Science} \textbf{\bibinfo{volume}{297}},
  \bibinfo{pages}{359} (\bibinfo{year}{2002}).

\bibitem[{\citenamefont{Baddeley et~al.}(1997)\citenamefont{Baddeley, Abbott,
  Booth, Sengpiel, Freeman, Wakeman, and Rolls}}]{baddeley}
\bibinfo{author}{\bibfnamefont{R.}~\bibnamefont{Baddeley}},
  \bibinfo{author}{\bibfnamefont{L.~F.} \bibnamefont{Abbott}},
  \bibinfo{author}{\bibfnamefont{M.~C.} \bibnamefont{Booth}},
  \bibinfo{author}{\bibfnamefont{F.}~\bibnamefont{Sengpiel}},
  \bibinfo{author}{\bibfnamefont{T.}~\bibnamefont{Freeman}},
  \bibinfo{author}{\bibfnamefont{E.~A.} \bibnamefont{Wakeman}},
  \bibnamefont{and} \bibinfo{author}{\bibfnamefont{E.~T.} \bibnamefont{Rolls}},
  \bibinfo{journal}{Proceedings of the Royal Society of London B: Biological
  Sciences} \textbf{\bibinfo{volume}{264}}, \bibinfo{pages}{1775}
  (\bibinfo{year}{1997}).

\bibitem[{\citenamefont{Froudarakis et~al.}(2014)\citenamefont{Froudarakis,
  Berens, Ecker, Cotton, Sinz, Yatsenko, Saggau, Bethge, and
  Tolias}}]{froudarakis}
\bibinfo{author}{\bibfnamefont{E.}~\bibnamefont{Froudarakis}},
  \bibinfo{author}{\bibfnamefont{P.}~\bibnamefont{Berens}},
  \bibinfo{author}{\bibfnamefont{A.~S.} \bibnamefont{Ecker}},
  \bibinfo{author}{\bibfnamefont{R.~J.} \bibnamefont{Cotton}},
  \bibinfo{author}{\bibfnamefont{F.~H.} \bibnamefont{Sinz}},
  \bibinfo{author}{\bibfnamefont{D.}~\bibnamefont{Yatsenko}},
  \bibinfo{author}{\bibfnamefont{P.}~\bibnamefont{Saggau}},
  \bibinfo{author}{\bibfnamefont{M.}~\bibnamefont{Bethge}}, \bibnamefont{and}
  \bibinfo{author}{\bibfnamefont{A.~S.} \bibnamefont{Tolias}},
  \bibinfo{journal}{Nature neuroscience} \textbf{\bibinfo{volume}{17}},
  \bibinfo{pages}{851} (\bibinfo{year}{2014}).

\bibitem[{\citenamefont{Rolls and Tovee}(1995)}]{rolls1995}
\bibinfo{author}{\bibfnamefont{E.~T.} \bibnamefont{Rolls}} \bibnamefont{and}
  \bibinfo{author}{\bibfnamefont{M.~J.} \bibnamefont{Tovee}},
  \bibinfo{journal}{Journal of Neurophysiology} \textbf{\bibinfo{volume}{73}},
  \bibinfo{pages}{713} (\bibinfo{year}{1995}).

\bibitem[{\citenamefont{Rust and DiCarlo}(2012)}]{rust}
\bibinfo{author}{\bibfnamefont{N.~C.} \bibnamefont{Rust}} \bibnamefont{and}
  \bibinfo{author}{\bibfnamefont{J.~J.} \bibnamefont{DiCarlo}},
  \bibinfo{journal}{The Journal of Neuroscience} \textbf{\bibinfo{volume}{32}},
  \bibinfo{pages}{10170} (\bibinfo{year}{2012}).

\bibitem[{\citenamefont{Vinje and Gallant}(2000)}]{vinje}
\bibinfo{author}{\bibfnamefont{W.~E.} \bibnamefont{Vinje}} \bibnamefont{and}
  \bibinfo{author}{\bibfnamefont{J.~L.} \bibnamefont{Gallant}},
  \bibinfo{journal}{Science} \textbf{\bibinfo{volume}{287}},
  \bibinfo{pages}{1273} (\bibinfo{year}{2000}).

\bibitem[{\citenamefont{Hahnloser et~al.}(2002)\citenamefont{Hahnloser,
  Kozhevnikov, and Fee}}]{hahnloser}
\bibinfo{author}{\bibfnamefont{R.}~\bibnamefont{Hahnloser}},
  \bibinfo{author}{\bibfnamefont{A.}~\bibnamefont{Kozhevnikov}},
  \bibnamefont{and} \bibinfo{author}{\bibfnamefont{M.}~\bibnamefont{Fee}},
  \bibinfo{journal}{Nature} \textbf{\bibinfo{volume}{419}}, \bibinfo{pages}{65}
  (\bibinfo{year}{2002}).

\bibitem[{\citenamefont{Fried et~al.}(1997)\citenamefont{Fried, MacDonald, and
  Wilson}}]{fried}
\bibinfo{author}{\bibfnamefont{I.}~\bibnamefont{Fried}},
  \bibinfo{author}{\bibfnamefont{K.~A.} \bibnamefont{MacDonald}},
  \bibnamefont{and} \bibinfo{author}{\bibfnamefont{C.~L.}
  \bibnamefont{Wilson}}, \bibinfo{journal}{Neuron}
  \textbf{\bibinfo{volume}{18}}, \bibinfo{pages}{753} (\bibinfo{year}{1997}).

\bibitem[{\citenamefont{Kreiman et~al.}(2000)\citenamefont{Kreiman, Koch, and
  Fried}}]{kreiman}
\bibinfo{author}{\bibfnamefont{G.}~\bibnamefont{Kreiman}},
  \bibinfo{author}{\bibfnamefont{C.}~\bibnamefont{Koch}}, \bibnamefont{and}
  \bibinfo{author}{\bibfnamefont{I.}~\bibnamefont{Fried}},
  \bibinfo{journal}{Nature neuroscience} \textbf{\bibinfo{volume}{3}},
  \bibinfo{pages}{946} (\bibinfo{year}{2000}).

\bibitem[{\citenamefont{Quian~Quiroga et~al.}(2005)\citenamefont{Quian~Quiroga,
  Reddy, Krieman, Koch, and Fried}}]{qq1}
\bibinfo{author}{\bibfnamefont{R.}~\bibnamefont{Quian~Quiroga}},
  \bibinfo{author}{\bibfnamefont{L.}~\bibnamefont{Reddy}},
  \bibinfo{author}{\bibfnamefont{G.}~\bibnamefont{Krieman}},
  \bibinfo{author}{\bibfnamefont{C.}~\bibnamefont{Koch}}, \bibnamefont{and}
  \bibinfo{author}{\bibfnamefont{I.}~\bibnamefont{Fried}},
  \bibinfo{journal}{Nature} \textbf{\bibinfo{volume}{435}},
  \bibinfo{pages}{1102} (\bibinfo{year}{2005}).

\bibitem[{\citenamefont{Quiroga et~al.}(2009)\citenamefont{Quiroga, Kraskov,
  Koch, and Fried}}]{qq.multi}
\bibinfo{author}{\bibfnamefont{R.~Q.} \bibnamefont{Quiroga}},
  \bibinfo{author}{\bibfnamefont{A.}~\bibnamefont{Kraskov}},
  \bibinfo{author}{\bibfnamefont{C.}~\bibnamefont{Koch}}, \bibnamefont{and}
  \bibinfo{author}{\bibfnamefont{I.}~\bibnamefont{Fried}},
  \bibinfo{journal}{Current Biology} \textbf{\bibinfo{volume}{19}},
  \bibinfo{pages}{1308} (\bibinfo{year}{2009}).

\bibitem[{\citenamefont{Magyar and Collins}(2015)}]{twopop}
\bibinfo{author}{\bibfnamefont{A.}~\bibnamefont{Magyar}} \bibnamefont{and}
  \bibinfo{author}{\bibfnamefont{J.}~\bibnamefont{Collins}},
  \bibinfo{journal}{Phys. Rev. E.} \textbf{\bibinfo{volume}{92}},
  \bibinfo{pages}{012712} (\bibinfo{year}{2015}).

\bibitem[{\citenamefont{Ison et~al.}(2011)\citenamefont{Ison, Mormann, Cerf,
  Koch, Fried, and Quian~Quiroga}}]{ison}
\bibinfo{author}{\bibfnamefont{M.~J.} \bibnamefont{Ison}},
  \bibinfo{author}{\bibfnamefont{F.}~\bibnamefont{Mormann}},
  \bibinfo{author}{\bibfnamefont{M.}~\bibnamefont{Cerf}},
  \bibinfo{author}{\bibfnamefont{C.}~\bibnamefont{Koch}},
  \bibinfo{author}{\bibfnamefont{I.}~\bibnamefont{Fried}}, \bibnamefont{and}
  \bibinfo{author}{\bibfnamefont{R.}~\bibnamefont{Quian~Quiroga}},
  \bibinfo{journal}{Journal of Neurophysiology} \textbf{\bibinfo{volume}{106}},
  \bibinfo{pages}{1713} (\bibinfo{year}{2011}).

\bibitem[{\citenamefont{Quian~Quiroga et~al.}(2007)\citenamefont{Quian~Quiroga,
  Reddy, Koch, and Fried}}]{qq.decoding}
\bibinfo{author}{\bibfnamefont{R.}~\bibnamefont{Quian~Quiroga}},
  \bibinfo{author}{\bibfnamefont{L.}~\bibnamefont{Reddy}},
  \bibinfo{author}{\bibfnamefont{C.}~\bibnamefont{Koch}}, \bibnamefont{and}
  \bibinfo{author}{\bibfnamefont{I.}~\bibnamefont{Fried}},
  \bibinfo{journal}{Journal of neurophysiology} \textbf{\bibinfo{volume}{98}},
  \bibinfo{pages}{1997} (\bibinfo{year}{2007}).

\bibitem[{\citenamefont{O'Keefe and Dostrovsky}(1971)}]{okeefe}
\bibinfo{author}{\bibfnamefont{J.}~\bibnamefont{O'Keefe}} \bibnamefont{and}
  \bibinfo{author}{\bibfnamefont{J.}~\bibnamefont{Dostrovsky}},
  \bibinfo{journal}{Brain research} \textbf{\bibinfo{volume}{34}},
  \bibinfo{pages}{171} (\bibinfo{year}{1971}).

\bibitem[{\citenamefont{Wilson and McNaughton}(1994)}]{wilson}
\bibinfo{author}{\bibfnamefont{M.~A.} \bibnamefont{Wilson}} \bibnamefont{and}
  \bibinfo{author}{\bibfnamefont{B.~L.} \bibnamefont{McNaughton}},
  \bibinfo{journal}{Science} \textbf{\bibinfo{volume}{265}},
  \bibinfo{pages}{676} (\bibinfo{year}{1994}).

\bibitem[{\citenamefont{Mormann et~al.}(2008)\citenamefont{Mormann, Kornblith,
  Quiroga, Kraskov, Cerf, Fried, and Koch}}]{mormann}
\bibinfo{author}{\bibfnamefont{F.}~\bibnamefont{Mormann}},
  \bibinfo{author}{\bibfnamefont{S.}~\bibnamefont{Kornblith}},
  \bibinfo{author}{\bibfnamefont{R.~Q.} \bibnamefont{Quiroga}},
  \bibinfo{author}{\bibfnamefont{A.}~\bibnamefont{Kraskov}},
  \bibinfo{author}{\bibfnamefont{M.}~\bibnamefont{Cerf}},
  \bibinfo{author}{\bibfnamefont{I.}~\bibnamefont{Fried}}, \bibnamefont{and}
  \bibinfo{author}{\bibfnamefont{C.}~\bibnamefont{Koch}}, \bibinfo{journal}{The
  Journal of Neuroscience} \textbf{\bibinfo{volume}{28}}, \bibinfo{pages}{8865}
  (\bibinfo{year}{2008}).

\bibitem[{\citenamefont{MacKay}(2003)}]{mackay}
\bibinfo{author}{\bibfnamefont{D.~J.} \bibnamefont{MacKay}},
  \emph{\bibinfo{title}{Information theory, inference and learning algorithms}}
  (\bibinfo{publisher}{Cambridge university press}, \bibinfo{year}{2003}).

\bibitem[{\citenamefont{Buzs{\'a}ki and Mizuseki}(2014)}]{buzsaki.2014}
\bibinfo{author}{\bibfnamefont{G.}~\bibnamefont{Buzs{\'a}ki}} \bibnamefont{and}
  \bibinfo{author}{\bibfnamefont{K.}~\bibnamefont{Mizuseki}},
  \bibinfo{journal}{Nature Reviews Neuroscience} \textbf{\bibinfo{volume}{15}},
  \bibinfo{pages}{264} (\bibinfo{year}{2014}).

\bibitem[{\citenamefont{Buzs{\'a}ki et~al.}(2015)}]{buzsaki.2015}
\bibinfo{author}{\bibfnamefont{G.}~\bibnamefont{Buzs{\'a}ki}}
  \bibnamefont{et~al.}, \bibinfo{journal}{science}
  \textbf{\bibinfo{volume}{347}}, \bibinfo{pages}{612} (\bibinfo{year}{2015}).

\bibitem[{\citenamefont{Rich et~al.}(2014)\citenamefont{Rich, Liaw, and
  Lee}}]{rich}
\bibinfo{author}{\bibfnamefont{P.~D.} \bibnamefont{Rich}},
  \bibinfo{author}{\bibfnamefont{H.-P.} \bibnamefont{Liaw}}, \bibnamefont{and}
  \bibinfo{author}{\bibfnamefont{A.~K.} \bibnamefont{Lee}},
  \bibinfo{journal}{Science} \textbf{\bibinfo{volume}{345}},
  \bibinfo{pages}{814} (\bibinfo{year}{2014}).

\bibitem[{\citenamefont{Quian~Quiroga}(2012)}]{qq.concept}
\bibinfo{author}{\bibfnamefont{R.}~\bibnamefont{Quian~Quiroga}},
  \bibinfo{journal}{Nat. Rev. Neurosci.} \textbf{\bibinfo{volume}{13}},
  \bibinfo{pages}{587} (\bibinfo{year}{2012}).

\bibitem[{\citenamefont{Barab{\'a}si and Albert}(1999)}]{barabasi}
\bibinfo{author}{\bibfnamefont{A.-L.} \bibnamefont{Barab{\'a}si}}
  \bibnamefont{and} \bibinfo{author}{\bibfnamefont{R.}~\bibnamefont{Albert}},
  \bibinfo{journal}{science} \textbf{\bibinfo{volume}{286}},
  \bibinfo{pages}{509} (\bibinfo{year}{1999}).

\bibitem[{\citenamefont{Newman}(2005)}]{newman}
\bibinfo{author}{\bibfnamefont{M.~E.} \bibnamefont{Newman}},
  \bibinfo{journal}{Contemporary physics} \textbf{\bibinfo{volume}{46}},
  \bibinfo{pages}{323} (\bibinfo{year}{2005}).

\bibitem[{\citenamefont{Mahmoud}(2008)}]{mahmoud}
\bibinfo{author}{\bibfnamefont{H.}~\bibnamefont{Mahmoud}},
  \emph{\bibinfo{title}{P{\'o}lya urn models}} (\bibinfo{publisher}{CRC press},
  \bibinfo{year}{2008}).

\bibitem[{\citenamefont{Peruani et~al.}(2007)\citenamefont{Peruani, Choudhury,
  Mukherjee, and Ganguly}}]{peruani}
\bibinfo{author}{\bibfnamefont{F.}~\bibnamefont{Peruani}},
  \bibinfo{author}{\bibfnamefont{M.}~\bibnamefont{Choudhury}},
  \bibinfo{author}{\bibfnamefont{A.}~\bibnamefont{Mukherjee}},
  \bibnamefont{and} \bibinfo{author}{\bibfnamefont{N.}~\bibnamefont{Ganguly}},
  \bibinfo{journal}{EPL (Europhysics Letters)} \textbf{\bibinfo{volume}{79}},
  \bibinfo{pages}{28001} (\bibinfo{year}{2007}).

\bibitem[{\citenamefont{Collins and Jin}(2006)}]{collins}
\bibinfo{author}{\bibfnamefont{J.}~\bibnamefont{Collins}} \bibnamefont{and}
  \bibinfo{author}{\bibfnamefont{D.~Z.} \bibnamefont{Jin}},
  \bibinfo{journal}{arXiv preprint q-bio/0603014}  (\bibinfo{year}{2006}).

\bibitem[{\citenamefont{Waydo et~al.}(2006)\citenamefont{Waydo, Kraskov,
  Quiroga, Fried, and Koch}}]{waydo}
\bibinfo{author}{\bibfnamefont{S.}~\bibnamefont{Waydo}},
  \bibinfo{author}{\bibfnamefont{A.}~\bibnamefont{Kraskov}},
  \bibinfo{author}{\bibfnamefont{R.~Q.} \bibnamefont{Quiroga}},
  \bibinfo{author}{\bibfnamefont{I.}~\bibnamefont{Fried}}, \bibnamefont{and}
  \bibinfo{author}{\bibfnamefont{C.}~\bibnamefont{Koch}}, \bibinfo{journal}{The
  Journal of Neuroscience} \textbf{\bibinfo{volume}{26}},
  \bibinfo{pages}{10232} (\bibinfo{year}{2006}).

\bibitem[{\citenamefont{Henze et~al.}(2000)\citenamefont{Henze, Borhegyi,
  Csicsvari, Mamiya, Harris, and Buzs{\'a}ki}}]{henze}
\bibinfo{author}{\bibfnamefont{D.~A.} \bibnamefont{Henze}},
  \bibinfo{author}{\bibfnamefont{Z.}~\bibnamefont{Borhegyi}},
  \bibinfo{author}{\bibfnamefont{J.}~\bibnamefont{Csicsvari}},
  \bibinfo{author}{\bibfnamefont{A.}~\bibnamefont{Mamiya}},
  \bibinfo{author}{\bibfnamefont{K.~D.} \bibnamefont{Harris}},
  \bibnamefont{and}
  \bibinfo{author}{\bibfnamefont{G.}~\bibnamefont{Buzs{\'a}ki}},
  \bibinfo{journal}{Journal of neurophysiology} \textbf{\bibinfo{volume}{84}},
  \bibinfo{pages}{390} (\bibinfo{year}{2000}).

\bibitem[{\citenamefont{Thompson and Best}(1989)}]{thompson}
\bibinfo{author}{\bibfnamefont{L.}~\bibnamefont{Thompson}} \bibnamefont{and}
  \bibinfo{author}{\bibfnamefont{P.}~\bibnamefont{Best}}, \bibinfo{journal}{The
  Journal of neuroscience} \textbf{\bibinfo{volume}{9}}, \bibinfo{pages}{2382}
  (\bibinfo{year}{1989}).

\bibitem[{\citenamefont{Olshausen and Field}(2005)}]{olshausen1}
\bibinfo{author}{\bibfnamefont{B.}~\bibnamefont{Olshausen}} \bibnamefont{and}
  \bibinfo{author}{\bibfnamefont{D.}~\bibnamefont{Field}}, pp.
  \bibinfo{pages}{1665--1699} (\bibinfo{year}{2005}).

\end{thebibliography}

\end{document}